\documentclass[a4paper,10pt]{IEEEtran}

\usepackage{amsmath}
\usepackage{amssymb}
\usepackage{amsfonts}
\usepackage{graphicx}
\usepackage{cite}

\usepackage{optidef}

\usepackage{xcolor}

\usepackage{physics}

\usepackage{setspace}

\usepackage{tabularx}

\newcommand\blfootnote[1]{%
  \begingroup
  \renewcommand\thefootnote{}\footnote{#1}%
  \addtocounter{footnote}{-1}%
  \endgroup
}

\title{Quantum Machine Learning for Distributed  Quantum Protocols with Local Operations and Noisy Classical Communications}

\author{
\IEEEauthorblockN{Hari Hara Suthan Chittoor,~\IEEEmembership{Member,~IEEE } and Osvaldo Simeone,~\IEEEmembership{Fellow,~IEEE }}\vspace{-0.2cm}}

\begin{document}

\maketitle

\thispagestyle{empty}	

\pagestyle{empty}



\blfootnote{The authors are with King’s Communications, Learning, and Information
Processing (KCLIP) lab at the Department of Engineering of Kings College
London, UK (emails: hari.hara@kcl.ac.uk, osvaldo.simeone@kcl.ac.uk).
The authors have received funding from
the European Research Council (ERC) under the European Union’s Horizon 2020 Research and Innovation Programme (grant agreement No. 725732), by the European Union’s Horizon Europe project CENTRIC (101096379), and by an Open Fellowship of the EPSRC (EP/W024101/1).

The PyTorch and MATLAB code for regenerating the results of this paper is available at $<$https://github.com/kclip/Noise-Aware-LOCCNet$>$.
}


\begin{abstract}
Distributed quantum information processing protocols such as quantum entanglement distillation and quantum state discrimination rely on local operations and classical communications (LOCC). Existing LOCC-based protocols typically assume the availability of ideal, noiseless, communication channels. In this paper, we study the case in which classical communication takes place over noisy channels, and we propose to address the design of LOCC protocols in this setting via the use of quantum machine learning tools. We specifically focus on the important tasks of quantum entanglement distillation and quantum state discrimination, and implement local processing through parameterized quantum circuits (PQCs) that are optimized to maximize the average fidelity and average success probability in the respective tasks, while accounting for communication errors.  The introduced approach, Noise Aware-LOCCNet (NA-LOCCNet), is shown to have significant advantages over existing protocols designed for noiseless communications.
\end{abstract}

\begin{IEEEkeywords}
Quantum machine learning, entanglement distillation, state discrimination, distributed quantum computing, parameterized quantum circuits
\end{IEEEkeywords}

\section{Introduction}

\subsection{Motivation}
Distributed quantum computing is considered to be an important application for the quantum Internet, offering a path forward towards scalable quantum computers \cite{IEEE_Network_quantum_internet_challenges_in_distributed_quantum_computing}. A practically and theoretically relevant class of distributed quantum computing protocols relies on \emph{local quantum operations and classical communications (LOCC)} \cite{Teleportation_Bennett,Everything_you_always_wanted_to_know_about_LOCC_(but_were_afraidtoask)}. In LOCC-based protocols,  distributed nodes carry out local quantum processing steps that are interwoven with the exchange of classical information, i.e., bits. LOCC-based protocols have been designed for a variety of tasks, including   entanglement distillation, state discrimination, channel simulation 
\cite{DISTILLATION_BBPSW, QSD_LOCC_fundamental_paper_2000,Fund_lim_of_repeaterless_Q_comm_pirandola_2017,Simulation_of_non_Pauli_channels}.

Recently, a \emph{quantum machine learning (QML)} framework was introduced in \cite{LOCCNet_Nature_2021} for the design of LOCC protocols. The approach, termed \textit{LOCCNet}, is motivated by the difficulty of designing optimal LOCC protocols under the restrictions imposed by noisy intermediate scale quantum (NISQ) computers. Following the QML framework \cite{book_machine_learning_with_quantum_computers_Maria_Francesco,Book_Osvaldo_quantum_machine_learning_for_engineers}, LOCCNet prescribes the use of \emph{parameterized quantum circuits (PQCs)} for local processing.  PQCs have been widely investigated in recent years as means to program NISQ computers via classical optimization, with applications ranging from combinatorial optimization to generative modelling \cite{book_machine_learning_with_quantum_computers_Maria_Francesco}. 
A PQC typically consists of a sequence of one- and two-qubit rotations, whose parameters can be optimized, as well as of fixed entangling gates.

Existing LOCC-based protocols, including the QML-based schemes introduced in \cite{LOCCNet_Nature_2021}, assume the availability of ideal, noiseless, communication channels. In contrast, in this paper, we study the case in which classical communication takes place over noisy channels. We introduce an approach, referred to as \emph{Noise Aware-LOCCNet (NA-LOCCNet)}, that addresses the design of LOCC protocols in the presence of noisy classical channels via the use of QML tools. We specifically focus on the important tasks of quantum entanglement distillation and quantum state discrimination, and implement local processing through PQCs that are optimized to maximize the average fidelity and average success probability in the respective tasks, while accounting for communication errors.  

\subsection{Entanglement distillation}


        
\begin{figure}[htbp]
\centering
\includegraphics[height=2.1in]{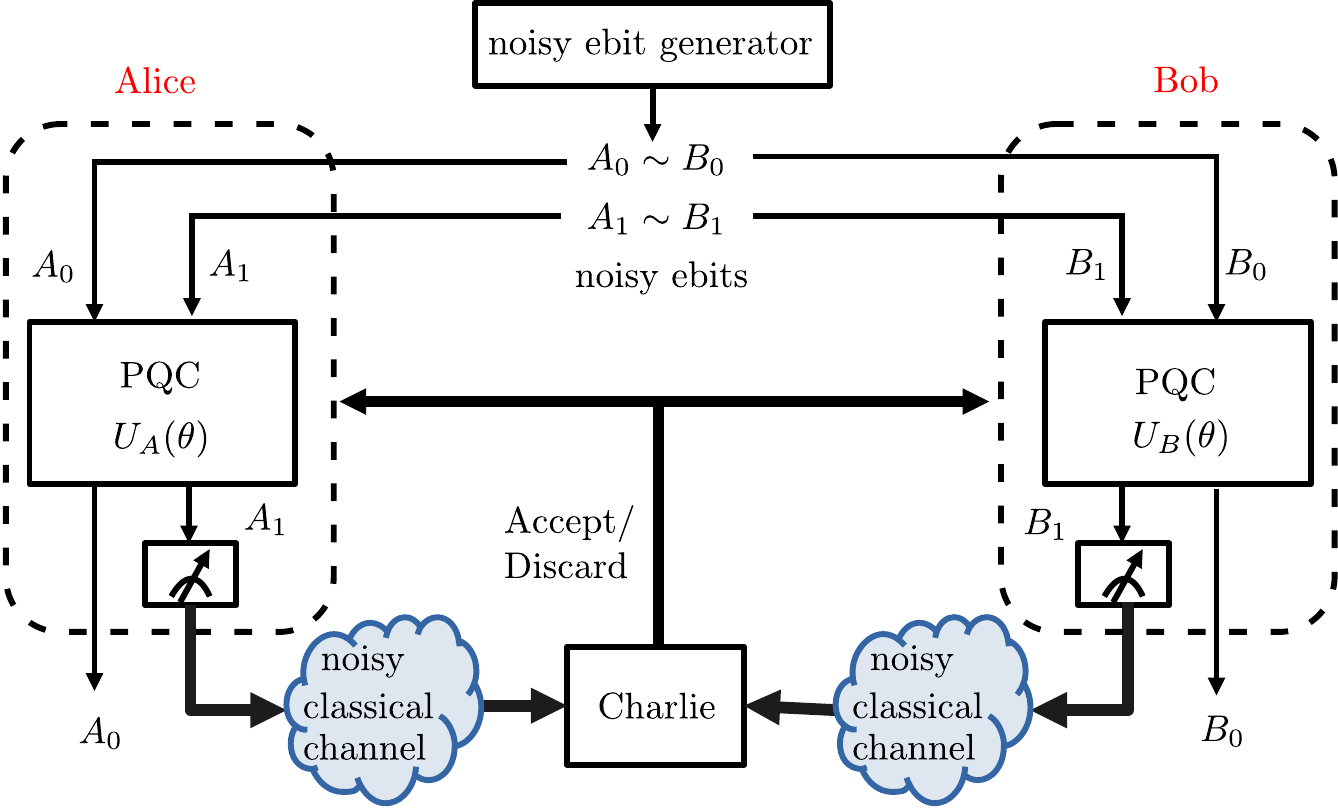} 
\caption{Distributed entanglement distillation at two quantum-enabled devices (Alice and Bob) aided by a noisy classical communication channel to a third party (Charlie). Alice and Bob implement PQCs as local operations and they communicate over a noisy classical link from Alice to Bob. }
\label{fig: Ent_dist_problem_formulation}
\end{figure}


Quantum networking, and with it the quantum Internet, rely on the management and exploitation of entanglement \cite{Book_Quantum_internet_second_quantum_revolution_2021,Book_quantum_networking_Meter,IEEE_Network_quantum_internet_challenges_in_distributed_quantum_computing}. In fact, entangled qubits enable fundamental quantum communication primitives such as teleportation and superdense coding \cite{Book_quantum_computing_quantum_information_Nielsen_chuang_2010,Book_Quantum_Information_Theory_wilde_2013}. Practical sources of entangled qubits, such as single-photon detection \cite{single_photon_sources_and_detectors,Sstate_ref_paper2}, are imperfect, producing mixed states with reduced fidelity as compared to ideal, fully entangled, Bell pairs. In order to enhance the fidelity of entangled qubits available at distributed parties, entanglement distillation protocols leverage LOCC. 

In entanglement distillation protocols, a source produces a number of imperfectly entangled qubit pairs. Each qubit of a pair is made available at one of two parties, conventionally referred to as Alice and Bob. The goal is to leverage LOCC to produce qubit pairs that have a higher degree of fidelity with respect to a fully entangled Bell pair. In the most typical case, Alice and Bob start with two qubit pairs, and output either one qubit pair or a declaration of failure at the end of the process (see Fig. \ref{fig: Ent_dist_problem_formulation}, Fig. \ref{fig: Ent_dist_LOCCNet} and Fig. \ref{fig: Ent_dist_proposed NA-LOCCNet}).

Traditionally, entanglement distillation protocols have been designed by hand, targeting specific mixed states as the input of the protocol \cite{DISTILLATION_BBPSW,Book_quantum_networking_Meter,DISTILLATION_DEJMPS}. Specific examples include the DEJMPS protocol, which targets the so-called S-state \cite{DISTILLATION_DEJMPS}. These methods rely on local operations via specific unitaries; on the measurement of one qubit at Alice and Bob; and on classical communication of the measurement outputs on a noiseless channel. Based on the measurement outputs, Alice and Bob decide whether to keep the unmeasured pair of qubits or to declare a distillation failure.

Recently, as illustrated in Fig. \ref{fig: Ent_dist_problem_formulation}, the LOCCNet framework introduced in \cite{LOCCNet_Nature_2021} for the design of LOCC protocols  prescribes the use of PQCs for the local unitaries applied by Alice and Bob. LOCCNet assumes ideal classical communications, while this paper studies the case in which communications between the parties holding imperfectly entangled qubits takes place over a noisy channel. To address this more challenging scenario, the proposed NA-LOCCNet method leverages to adapt QML tools to program local operations via PQCs while accounting for channel noise.

\subsection{Quantum state discrimination}


\begin{figure}[htbp]
    \centering
    \includegraphics[height=3.5in]{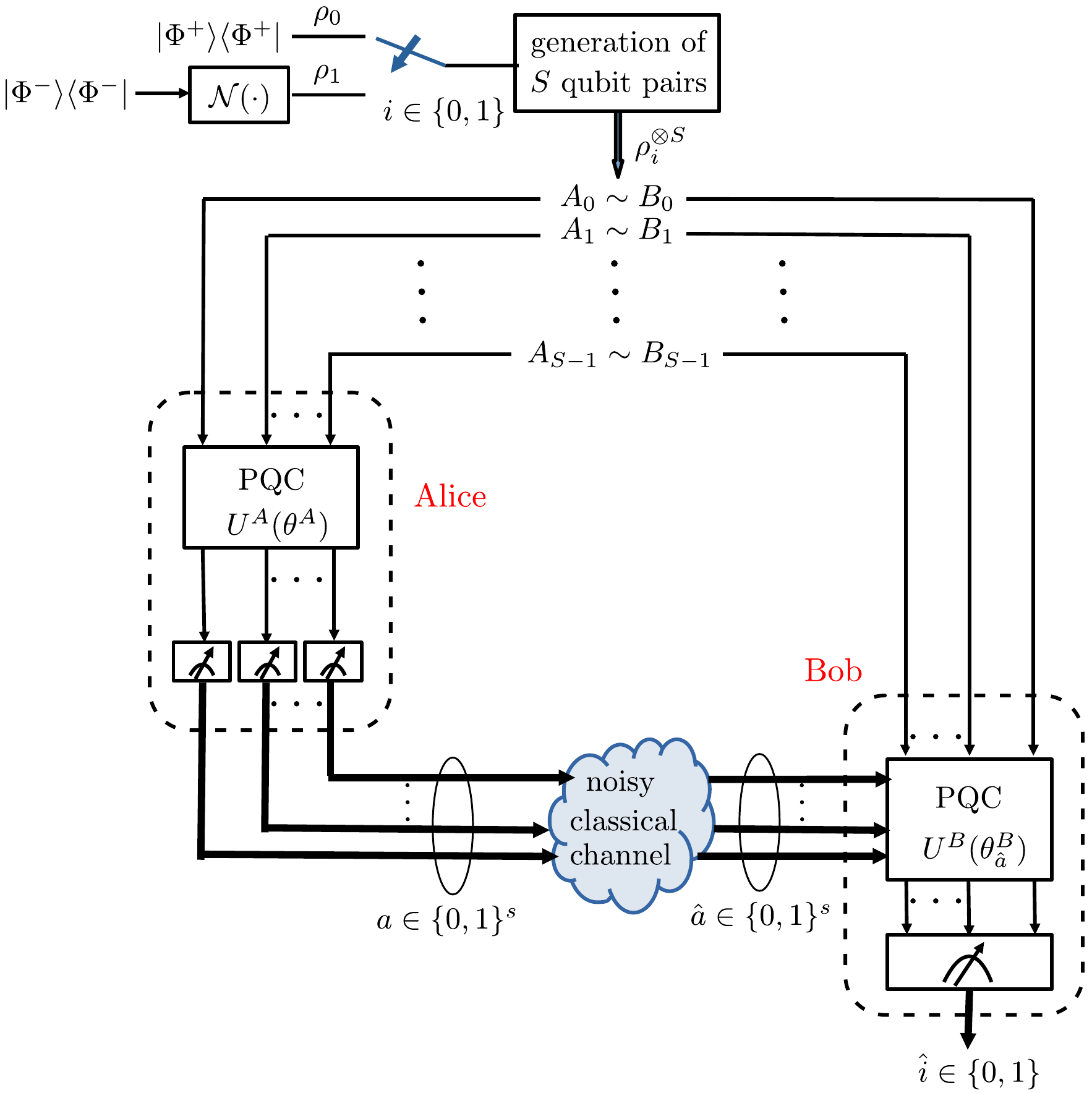}
    \caption{Distributed quantum state discrimination at two quantum-enabled devices, Alice and Bob. Alice and Bob implement parameterized quantum circuits (PQCs) as local operations and they communicate over a noisy classical link from Alice to Bob. }
    \label{fig: QSD problem formulation}
\end{figure}


Another important quantum information processing protocol is quantum state discrimination which is central to many applications of quantum sensing, communication, networking, and computing \cite{QSD_applications,metrology,cryptography}. Of particular interest for quantum sensing are settings in which distributed nodes have access to correlated quantum subsystems, and they are tasked with discriminating between two possible joint states of the overall system \cite{distributed_quantum_sensing}. As illustrated in Fig. \ref{fig: QSD problem formulation}, in such a situation, the nodes may be able to implement local operations  on their shares of the quantum system, as well as to communicate using classical communication links.

Traditionally, quantum state discrimination protocols based on LOCC protocols have been designed by hand by focusing on the discrimination of specific pairs of states. Specific examples include the discrimination of orthogonal pure states \cite{QSD_LOCC_fundamental_paper_2000} and the discrimination of maximally entangled states \cite{QSD_of_maximally_entangled_states}. Assuming the presence of two nodes, Alice and Bob, these methods select the unitary at Bob as a function of the output of measurements made by Alice and shared on a noiseless communication link with Bob.

Reference
\cite{LOCCNet_Nature_2021} also introduced the LOCCNet framework for quantum state discrimination.
The design of LOCCNet in \cite{LOCCNet_Nature_2021} considers the problem of distinguishing two orthogonal maximally entangled Bell states, where one of the Bell state is corrupted by an entanglement-breaking quantum channel. The design assumes ideal, noiseless, classical communications, and it operates on a single pair of qubits. As a second contribution, this paper introduces the NA-LOCCNet framework for quantum state discrimination by accounting for noisy classical communications in the design problem.  

\subsection{Main contributions}

As summarized in the previous subsections, the design of LOCCNet in \cite{LOCCNet_Nature_2021} assumes ideal, noiseless, classical communications. In contrast, in this paper, we study the case in which communication  takes place over \textit{noisy binary symmetric channels}. The specific contributions are as follows. 

\begin{itemize}
    \item  As seen in Fig. \ref{fig: Ent_dist_problem_formulation},  we first introduce NA-LOCCNet as a novel PQC-based architecture for distributed entanglement distillation (see Fig. \ref{fig: Ent_dist_proposed NA-LOCCNet}) that is designed with the goal of maximizing the average fidelity while accounting for the randomness caused by communication errors.   

    \item  Then,  we adapt the NA-LOCCNet framework for the problem of distributed quantum state discrimination (see Fig. \ref{fig: QSD two pair})  with the goal of maximizing the average probability of successful detection for quantum state discrimination.  

    \item The introduced NA-LOCCNet is shown via experiments to have significant advantages over existing protocols designed for noiseless communications. Furthermore, in quantum state discrimination we make the important observation that, depending on the level of classical noise, a larger level of entanglement-breaking noise can be advantageous to facilitate successful  distributed discrimination.
\end{itemize}



Part of this paper was presented in \cite{Hari_osvaldo_entanglement_distillation_LOCCNet}, which  covered only NA-LOCCNet for entanglement distillation.

\subsection{Organization}
The rest of the paper is organized as follows. In Section \ref{section Learning Entanglement Distillation} we present NA-LOCCNet protocol for distributed entanglement distillation, while Section \ref{section Learning Quantum State Discrimination} focuses on the NA-LOCCNet protocol for distributed quantum state discrimination. In both sections,  we first define the problem statement, review the relevant state of the art, present the proposed NA-LOCCNet protocol, and finally give experimental results\footnote{The experiments in Section $2.4$ and Section $3.4$ are carried out on a quantum simulator run on a laptop with i7 processor and 16 GB RAM. We did not consider the impact of quantum hardware noise \cite{Practical_QEM_for_Near_Future_Applications, Error_Correcting_Codes_in_Quantum_Theory,Wood_noise_characterization_error_mitigation_NISQ}.}. Section \ref{section conclusions} concludes the paper.


\subsection{Notations and definitions} 

For any non-negative integer $K$, $[K]$ represents the set $\{0,1,\cdots,K\}$. Given a discrete set $\mathcal{A}$ and positive integer $S$, $\mathcal{A}^S$ represents the set of strings of length $S$ from the alphabet $\mathcal{A}$. The Kronecker product is denoted as $\otimes$; $I_{d}$ represents the $d\times d$ identity matrix; $M^{\dagger}$ represents the complex conjugate transpose of the matrix $M$; $\mathrm{tr}(M)$ represents trace of the matrix $M$; and a positive semidefinite matrix $M$ is denoted as $M \succeq 0$. We adopt standard notations for quantum states, computational basis, and quantum gates \cite{Book_quantum_computing_quantum_information_Nielsen_chuang_2010}. Let $\mathcal{A}$ and $\mathcal{B}$ be two Hilbert spaces of dimensions $d^A$ and $d^B$, with computational basis $\{|i\rangle\}_{i=0}^{2^{d_A}-1}$ and $\{|j\rangle\}_{j=0}^{2^{d_B}-1}$, respectively. Any $2^{d_A + d_B} \times 2^{d_A + d_B}$ complex matrix $M$ on the Hilbert space $\mathcal{A}\otimes \mathcal{B}$, can be written as $M = \sum_{ijkl}p_{kl}^{ij} |i\rangle \langle j| \otimes |k\rangle \langle l|$, where $p_{kl}^{ij}$ are complex numbers and the sums range over the sets $i,j \in [2^{d_A -1}]$ and $k,l \in [2^{d_B -1}]$. The \textit{partial transpose operator} of $M$ with respect to $\mathcal{B}$ is defined as $M^{T_B}  =  \sum_{ijkl}p_{kl}^{ij} |i\rangle \langle j| \otimes |l\rangle \langle k|$. The \textit{partial trace} of $M$ with respect to $\mathcal{A}$ is defined as $\mathrm{tr}_{A}(M) = \sum_{i=0}^{2^{d_A}-1}(\langle i| \otimes I^{\mathcal{B}}) M (|i \rangle \otimes I^{\mathcal{B}})$, where $I^{\mathcal{B}}$ is the $2^{d_B} \times 2^{d_B}$ identity matrix.

\section{Learning Entanglement Distillation with Noisy Classical Communication}
\label{section Learning Entanglement Distillation}

In this section we first formulate the distributed entanglement distillation problem and review the relevant state of the art protocols. We then propose NA-LOCCNet for distributed entanglement distillation and give experimental results.

\subsection{Problem formulation}
\label{subsection problem formulation in entanglement distillation}

In this subsection, we formulate the problem of distributed entanglement distillation in the presence of a noisy classical communication channel, and we describe the performance metrics of interest.


\subsubsection{Setting}
\label{subsection setting in entanglement distillation}

As illustrated in Fig. \ref{fig: Ent_dist_problem_formulation}, we consider a system consisting of two main parties -- Alice and Bob -- aided by a third party -- Charlie. Alice and Bob have local quantum processing capability, while Charlie is not equipped with quantum computing devices. Alice and Bob can communicate to Charlie over a \textit{noisy classical channel}. An imperfect quantum entanglement mechanism generates pairs of noisy entangled qubits, also referred to as \textit{noisy ebits}. One of the qubits of each entangled pair is made available to Alice and the other to Bob. The goal of the system is to improve the average fidelity, defined in Section \ref{subsubsection performance metrics in entanglement distillation} and Section \ref{subsubsection design objective in entanglement distillation}, of the noisy ebits shared by Alice and Bob through local operations (LO) at Alice and Bob, as well as through classical communication (CC) to Charlie.

The quantum entanglement generator produces $k$ pairs of noisy ebits. The state of each qubit pair is described by a $4 \times 4$ density matrix $\rho_{AB}$. Throughout the paper, we use subscript $A$ to denote the qubits available at Alice, while the subscript $B$ is used for the qubits at Bob. As in \cite{LOCCNet_Nature_2021}, we specifically focus on the noisy, i.e., mixed, ebit state described by the density matrix
\begin{equation}
\label{eq: S-state}    
    \rho_{A B} = F | \phi^+ \rangle \langle \phi^+ | + (1-F) |00\rangle  \langle 00|,
\end{equation}
where $F \in [0,1]$ represents the \textit{input fidelity} and 
\begin{equation}
\label{eq: phi+ ebit}
    | \phi^+ \rangle = \frac{1}{\sqrt{2}} (|00\rangle + |11\rangle) 
\end{equation}
is a maximally entangled Bell state. The noisy ebit state in (\ref{eq: S-state}) is also known as \textit{S-state} \cite{LOCCNet_Nature_2021}, and it describes a situation in which the two qubits are in the maximally entangled state, $| \phi^+\rangle$, with probability $F$, and in the separable, i.e., non-entangled, state $|00 \rangle$ with probability $1-F$. This type of noisy state arise in the protocols for entanglement generation that use single-photon detection in the presence of photon loss \cite{Optimizing_practical_entanglement_distillation,Sstate_ref_paper1,Sstate_ref_paper2}. Furthermore, the S state is known to be more challenging to "denoise" than other mixed states in which the separable state, occurring with probability $1-F$, is orthogonal to $|\phi^+\rangle$ \cite{Optimizing_practical_entanglement_distillation}.

As in \cite{LOCCNet_Nature_2021}, we focus on the standard case in which $k=2$ identical pairs of S-states $\rho_{A_0 B_0}$ and $\rho_{A_1 B_1}$ are generated. The goal is to \textit{distill} the two noisy ebits pairs to obtain a single pair of less noisy ebits. Following standard terminology \cite{Book_quantum_networking_Meter}, the qubits $A_0$ and $B_0$ are referred to as the \textit{preserved pair}, and the qubits $A_1$ and $B_1$ as the \textit{sacrificial pair}. As shown in Fig. \ref{fig: Ent_dist_problem_formulation}, Alice and Bob process the respective qubits -- $A_1$ and $A_0$ for Alice, and $B_1$ and $B_0$ for Bob -- via \textit{local quantum operations} defined by unitaries $U_A(\theta)$ and $U_B(\theta)$ respectively. As detailed in the next sections, the operation of the unitaries generally depend on a vector $\theta$ of classical parameters. Then, the qubits $A_1$ and $B_1$ are measured in the computational basis at Alice and Bob respectively, and the measurement outcomes ($0$ or $1$) are communicated to Charlie using noisy classical channels. We specifically assume that communication to Charlie occurs over independent \textit{binary symmetric channels} with bit flip probability $p$.

If Charlie receives message $0$ from both Alice and Bob, it declares that the distillation is successful, and Alice and Bob retain the pair of qubits $A_0$ and $B_0$. Instead, if Charlie receives the pairs of messages $(0,1),(1,0)$ or $(1,1)$ from Alice and Bob, it declares a failure. In this case, Alice and Bob discard the qubits $A_0$ and $B_0$. 

We remark that most conventional entanglement distillation protocols \cite{DISTILLATION_BBPSW,DISTILLATION_DEJMPS} use decision rules in which either pair of messages $(0,0)$ or $(1,1)$ is considered as success. Here we follow the approach in \cite{LOCCNet_Nature_2021} of treating $(0,0)$ as the only case in which Charlie declares success. This design choice facilitates the optimization of the unitaries $U_A(\theta)$ and $U_B(\theta)$ through vector $\theta$.

One of the goals of this work is to design the unitaries $U_A(\theta)$ and $U_B(\theta)$ at Alice and Bob such that the output state of qubits $A_0$ and $B_0$, upon successful distillation, is as close as possible in terms of fidelity to the ideal ebit state $|\phi^+ \rangle$.


\subsubsection{Performance metrics}
\label{subsubsection performance metrics in entanglement distillation}

The performance of entanglement distillation is measured in this paper, as in \cite{LOCCNet_Nature_2021,Entanglement_purification_review}, in terms of fidelity and probability of success. The \textit{fidelity} of a state $\rho_{AB}$ with respect to the ebit state $|\phi^+\rangle$ is defined as \vspace{-0.2cm}
\begin{equation}
\label{eq: fidelity definition}
    F(\rho_{AB}) = \langle \phi^+| \rho_{AB} |\phi^+ \rangle ,    
\end{equation}
while \textit{probability of success} is the probability of receiving the pair of messages $(0,0)$ at Charlie.

Let $U(\theta)$ be the $16\times 16$ unitary operation corresponding to the separate application of the $4 \times 4$ local unitaries $U_A(\theta)$ and $U_B(\theta)$ to their respective qubit pairs $(A_0,A_1)$ and $(B_0,B_1)$, respectively. We order the qubits as $(A_0,B_0,A_1,B_1)$ to facilitate the derivations below. The state of the four qubits after the local operations can be expressed as the density matrix
\begin{equation}
\label{eq: rho_out}
    \rho_{out}(\theta) = U(\theta)  (\rho_{A_0 B_0} \otimes \rho_{A_1 B_1})  U(\theta)^{\dagger},
\end{equation}
where we have made explicit dependence on the model parameter vector $\theta$.

The measurement of the sacrificial pair of qubits $(A_1,B_1)$ in the computational basis, $\left\{|00\rangle, |01\rangle, |10\rangle, |11\rangle \right\}$, consists of the projective measurement defined by the four projection matrices
\begin{equation}
\label{eq: POVM}
    \Pi^{xy} =  I_4 \otimes |xy \rangle \langle xy| ,  
\end{equation}
with $(x,y) \in \{0,1\}^2$, where $I_4$ is the $4 \times 4$ identity matrix. Accordingly, the measurement returns output $(x,y) \in \{0,1\}^2$ with probability \vspace{-0.2cm}
\begin{equation}
\label{eq: postmeasurement state probability}
    P^{xy} (\theta) = \mathrm{tr}(\Pi^{xy} \rho_{out}(\theta) ),
\end{equation}
and the corresponding post-measurement state for the qubits $(A_0,B_0)$ is \vspace{-0.2cm}
\small
\begin{equation}
\label{eq: postmeasurement state}
    \rho_{A_0 B_0}^{xy}(\theta) = \frac{(I_4 \otimes \langle xy| ) \rho_{out}(\theta) (I_4 \otimes |xy \rangle) }{P^{xy}(\theta) }.
\end{equation}
\normalsize
Conditioned on the measurement outcome being $(x,y)\in \{0,1\}^2$, the fidelity (\ref{eq: fidelity definition}) of the state $\rho_{A_0 B_0}^{xy}(\theta)$ with respect to the ebit state $|\phi^+\rangle$ is hence
\begin{equation}
\label{eq: fidelity of postmeasurement state}
    F^{xy}(\theta) = \langle \phi^+| \rho_{A_0 B_0}^{xy}(\theta) |\phi^+ \rangle .
\end{equation}


\subsection{Existing distillation protocols}
\label{subsection literature review in entanglement distillation}

In this section, we review current state-of-the-art distillation protocols. We focus on the DEJMPS protocol \cite{DISTILLATION_DEJMPS} and on the LOCCNet protocol \cite{LOCCNet_Nature_2021} as applied to $k=2$ copies of the S-state (\ref{eq: S-state}). We emphasize that all the existing distillation protocols are designed for noiseless classical communication channels to Charlie, i.e., assuming $p=0$.

\subsubsection{DEJMPS protocol}
\label{subsubsection DEJMPS protocols}

In the DEJMPS protocol, the local unitaries $U_A(\theta)$ and $U_B(\theta)$ applied by Alice and Bob do not have free parameters, and are hence denoted as $U_A$ and $U_B$, dropping the dependence on the model parameter vector $\theta$. Specifically, the unitary $U_A$ at Alice is given by Pauli $X$-rotation 
$R_X(\pi/2)$ applied on both qubits, followed by a controlled NOT (CNOT) gate with the qubit $A_0$ as control and the qubit $A_1$ as target. Similarly, the unitary $U_B$ at Bob is defined by the cascade of Pauli $X$-rotations $R_X(-\pi/2)$ on the two qubits and of a CNOT gate with the qubit $B_0$ as control and the qubit $B_1$ as target. If Charlie receives messages $(0,0)$ or $(1,1)$ from Alice and Bob, it declares that distillation is successful, and the qubit pair $(A_0,B_0)$ is retained.

        
\begin{figure}[htbp]
\centering
\includegraphics[height=2.2in]{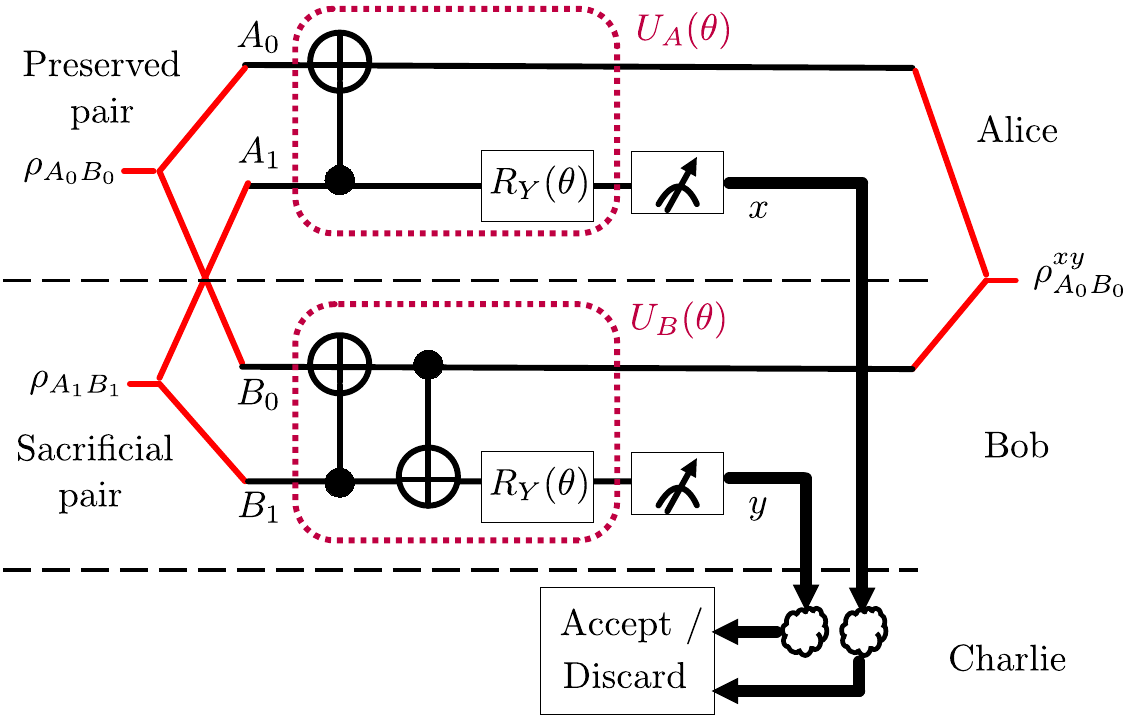}
\caption{LOCCNet circuit for distributed entanglement distillation of two S states \cite{LOCCNet_Nature_2021}.}
\label{fig: Ent_dist_LOCCNet}
\end{figure}


\vspace{-.3cm}
\subsubsection{LOCCNet}
\label{subsubsection LOCCNet in entanglement distillation}

In \cite{LOCCNet_Nature_2021}, a quantum machine learning-based entanglement distillation protocol, known as LOCCNet, is introduced that uses parameterized quantum circuits (PQCs) for unitaries $U_A(\theta)$ and $U_B(\theta)$ at Alice and Bob. As illustrated in Fig. \ref{fig: Ent_dist_LOCCNet}, the PQC $U_A(\theta)$ consists of a CNOT gate followed by a Pauli $Y$-rotation; while the PQC $U_B(\theta)$ is given by two CNOT gates followed by a Pauli $Y$-rotation. The rotation angle $\theta$ of the Pauli $Y$-rotation is subject to optimization. If Charlie receives messages $(0,0)$ from Alice and Bob through noiseless channels, i.e., $p=0$, a success is declared and the pair $(A_0,B_0)$ of qubits is retained. Model parameter vector $\theta$ is optimized with the goal of maximizing the fidelity $F^{00}(\theta)$ in (\ref{eq: fidelity of postmeasurement state}).


\subsection{Noise Aware-LOCCNet}
\label{subsection Noisy-LOCCNet in entanglement distillation}

In this section, we propose \textit{Noise Aware-LOCCNet} (NA-LOCCNet), which distills two qubit pairs, each in the S-state (\ref{eq: S-state}), in the presence of noisy classical channels from Alice and Bob to Charlie as shown in Fig. \ref{fig: Ent_dist_problem_formulation}. The key innovation as compared to LOCCNet is that we explicitly target the performance in terms of average fidelity by accounting for the impact of channel errors. We first describe the design objective, and then introduce the assumed structure for the PQCs $U_A(\theta)$ and $U_B(\theta)$.


\subsubsection{Design objective}
\label{subsubsection design objective in entanglement distillation}

NA-LOCCNet aims at maximizing the \textit{average conditional fidelity} of a retained pair $(A_0,B_0)$ in case of success. As explained in Section \ref{subsection setting in entanglement distillation}, Charlie declares a success if it receives the pair of messages $(0,0)$ from Alice and Bob through the respective binary symmetric channels with bit flip probability $p$. LOCCNet assumes a noiseless channel $(p=0)$, and hence it targets the objective $F^{00}(\theta)$, that is, the fidelity conditioned on measurement $(0,0)$ being produced by Alice and Bob. In contrast, NA-LOCCNet accounts for the fact that, where Charlie declares a success as it receives messages $(0,0)$, the actual measurement outcomes may be different due to channel errors.

In fact, messages $(0,0)$ are received at Charlie with probability $\mathcal{P}^{00} = (1-p)^2$ if the measurement outcomes are $(x,y) = (0,0)$; with probability $\mathcal{P}^{01} = (1-p)p$ if the measurement outcomes are $(x,y) = (0,1)$; with probability $\mathcal{P}^{10} = p(1-p)$ if the measurement outcomes are $(x,y) = (1,0)$; and
with probability $\mathcal{P}^{11} = p^2$ if the measurement outcomes are $(x,y) = (1,1)$.
Therefore, the average fidelity conditioned on the reception of messages $(0,0)$ is computed as

\small
\begin{align} 
    \label{eq: average fidelity for 00 success}
    \overline{F} (\theta) &=\frac{ \sum_{x,y}\mathcal{P}^{xy} P^{xy}(\theta) F^{xy}(\theta) }{P_{succ}(\theta)},
\end{align}
where 
\begin{align}
    \label{eq: average success probability for 00 case}
    P_{succ}(\theta) &= \sum_{x,y}\mathcal{P}^{xy} P^{xy}(\theta) 
\end{align} 
\normalsize
is the probability of success, i.e., of receiving messages $(0,0)$, and we have used definitions (\ref{eq: postmeasurement state probability}) and (\ref{eq: fidelity of postmeasurement state}). The proposed protocol NA-LOCCNet addresses the problem
\begin{equation}
\label{eq: optimization expression}
    \underset{\theta}{\max} ~ \overline{F}(\theta).    
\end{equation}

        
\begin{figure}[htbp]
\centering
\includegraphics[height=2.19in]{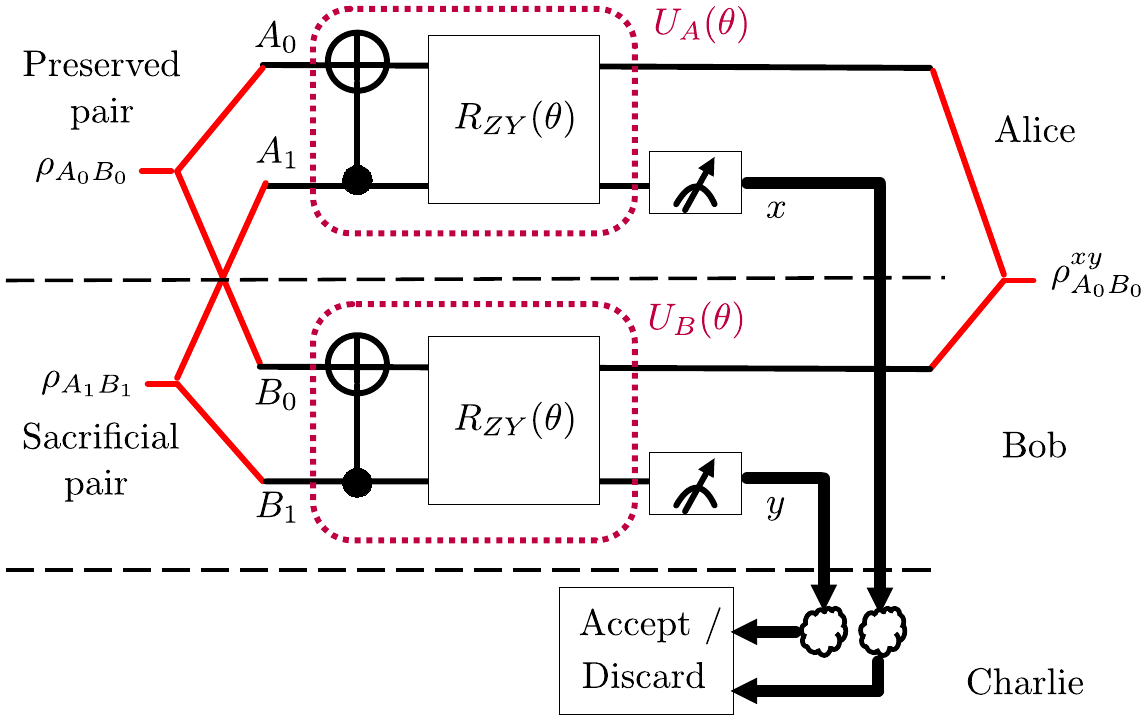}
\caption{Proposed Noise Aware-LOCCNet (NA-LOCCNet) circuit for distributed entanglement distillation of two S states.}
\label{fig: Ent_dist_proposed NA-LOCCNet}
\end{figure}



\subsubsection{Architecture of the PQCs}
\label{subsubsectin architecture of the PQCs in entanglement distillation}

For the PQCs $U_A(\theta)$ and $U_B(\theta)$ at Alice and Bob, respectively, we adopt the architecture shown in Fig. \ref{fig: Ent_dist_proposed NA-LOCCNet}. Unlike the LOCCNet architecture in Fig. \ref{fig: Ent_dist_LOCCNet}, we introduce a parameterized two-qubit gate, namely the Pauli $ZY$-rotation \cite{Ansatz_TWO_QUBIT_ROTAION_GATES}. This is defined by the unitary
\begin{equation}
\label{eq: two qubit ZY rotation gate}
    R_{ZY}(\theta) = \mathrm{exp}\left(-i \frac{\theta}{2} (Z \otimes Y)\right),
\end{equation}
which is parameterized by angle $\theta$. Recently, two-qubit rotation gates \cite{Ansatz_TWO_QUBIT_ROTAION_GATES} were shown to provide performance advantages as gates in PQCs for various quantum machine learning applications. In our work, the choice of the parameterized two-qubit gate (\ref{eq: two qubit ZY rotation gate}) was dictated by extensive experiments with alternative architectures\footnote{We tried various other ansatzes with different two qubit and single qubit rotation gates, changing the position of CNOT gate before and after the rotation gates, and changing the control and target qubits of CNOT gates. We note that the proposed ansatz in Fig. \ref{fig: Ent_dist_proposed NA-LOCCNet} gives best performance among the ansatzes we considered. The proposed architecture has the same complexity in terms of number of parameters as that of LOCCNet \cite{LOCCNet_Nature_2021}. We note that one could also consider ansatzes with more rotation angles for single qubit and two qubit rotation gates at Alice and Bob, and we leave an investigation of this point to future work.}. As an example, in Section \ref{subsection experiments in entanglement distillation}, we will compare the performance obtained by the architecture in Fig. \ref{fig: Ent_dist_proposed NA-LOCCNet} with the original LOCCNet system in Fig. \ref{fig: Ent_dist_LOCCNet}, when addressing problem (\ref{eq: optimization expression}).


\subsubsection{Optimization}
\label{subsubsection optimization in entanglement distillation}

Addressing problem (\ref{eq: optimization expression}) using QML with PQCs characterized by a single scalar parameter $\theta$, as for the architectures in Fig. \ref{fig: Ent_dist_LOCCNet} and Fig. \ref{fig: Ent_dist_proposed NA-LOCCNet}, requires a one-dimensional search over the limited domain $[0,2\pi)$. This can be carried out using standard optimization techniques, including grid search or gradient descent.
In particular, we use Adam gradient descent optimizer \cite{Adam_optimzer_fundamental_paper} with $0.01$ learning rate and $1001$ iterations. 
Similar to the vast majority of papers on quantum machine learning (see, e.g., \cite{LOCCNet_Nature_2021, PQC_benedetti2019parameterized}), the optimization is at the level of parameters, here $\theta$, of quantum gates. Implementation on a quantum computer requires a compilation step that accounts for the physical realization of the specific hardware \cite{Quantum_circuit_compilation_survey}.

\subsection{Experiments}
\label{subsection experiments in entanglement distillation}

In this section, we evaluate the performance of the proposed NA-LOCCNet protocol in the presence of noisy communication channels from Alice and Bob to Charlie.
We consider the benchmark schemes DEJMPS (Section \ref{subsubsection DEJMPS protocols}) and LOCCNet (Section \ref{subsubsection LOCCNet in entanglement distillation}). For the latter, we consider two designs: the original optimization in \cite{LOCCNet_Nature_2021} of the fidelity $F^{00}(\theta)$ in (\ref{eq: fidelity of postmeasurement state}) and the optimization of the conditional average fidelity $\overline{F}(\theta)$ in (\ref{eq: average fidelity for 00 success}) for the PQC architecture in Fig. \ref{fig: Ent_dist_LOCCNet}.



\begin{figure}[htbp]
    \centering
    \includegraphics[height=2.7in]{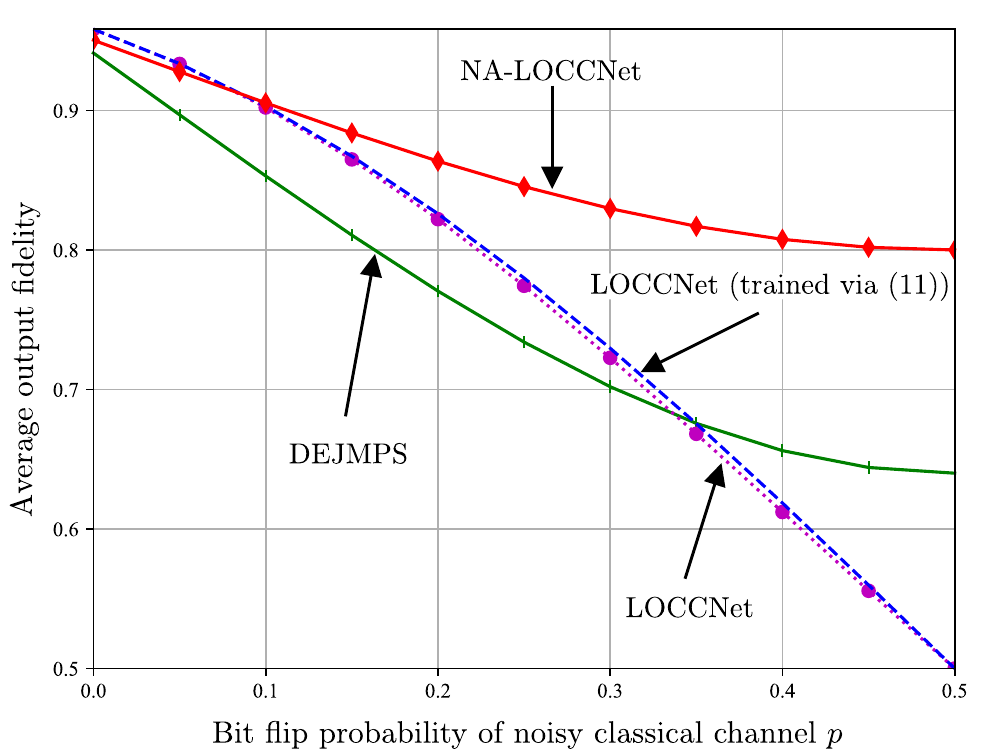}  
    \caption{Average output fidelity as a function of the bit flip probability $p$ of the noisy classical channels from Alice and Bob to Charlie for input fidelity $F=0.6$ in (\ref{eq: S-state}).}
    \label{fig: Ent_dist_Figure_output_fidelity_vs_bit_flip_probability.pdf}
\end{figure}



\begin{figure}[htbp]
    \centering
    \includegraphics[height=2.7in]{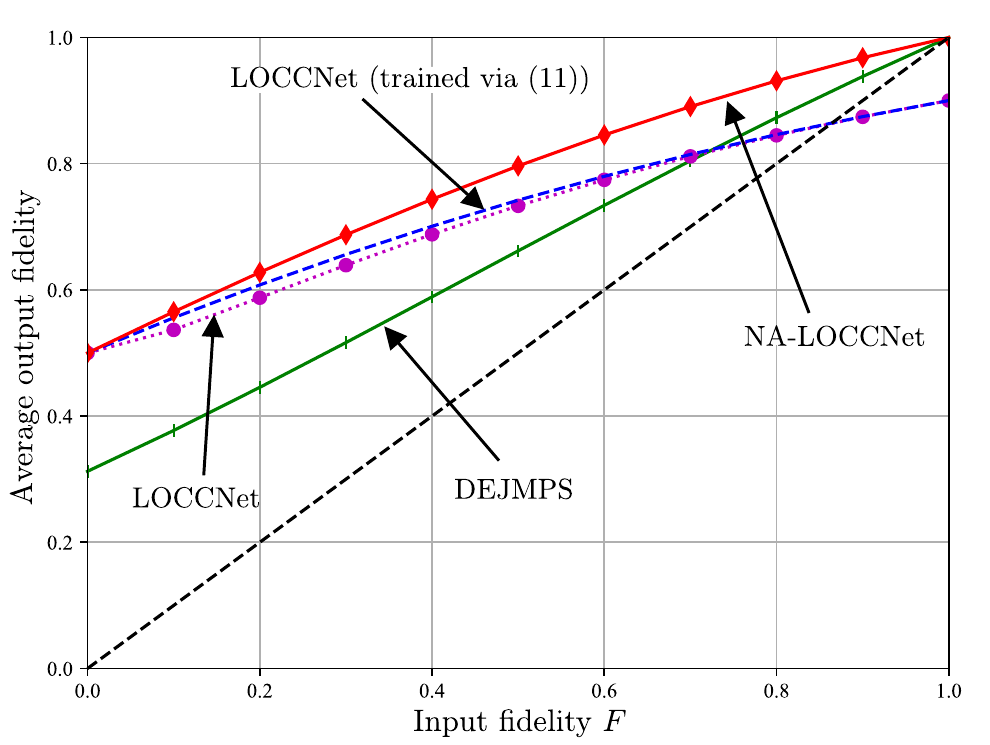}  
    \caption{Average output fidelity, conditioned on a successful distillation, as a function of the input fidelity $F$ in (\ref{eq: S-state}) for bit flip probability $p=0.25$ on the noisy classical channels from Alice and Bob to Charlie. The black dashed line corresponds to the reference performance of a scheme that simply outputs the input state.}
    \label{fig: Ent_dist_Figure_output_fidelity_vs_input_fidelity.pdf}
\end{figure}


Fig. \ref{fig: Ent_dist_Figure_output_fidelity_vs_bit_flip_probability.pdf} plots the average output fidelity, conditioned on a successful distillation, as a function of the bit flip probability $p$ of the noisy classical channels by fixing the input fidelity of the S-state (\ref{eq: S-state}) to $F=0.6$; while Fig. \ref{fig: Ent_dist_Figure_output_fidelity_vs_input_fidelity.pdf} plots the same quantity as a function of the input fidelity $F$ by fixing the bit flip probability to $p=0.25$. Note that the conditional average fidelity is given by (\ref{eq: average fidelity for 00 success}) for LOCCNet and NA-LOCCNet, while for DEJMPS one needs to consider both received messages $(0,0)$ and $(1,1)$ as indicating successful distillation.


Fig. \ref{fig: Ent_dist_Figure_output_fidelity_vs_bit_flip_probability.pdf} shows that, as the bit flip probability $p$ increases, the average fidelity of both DEJMPS and LOCCNet decreases significantly, reaching the minimum fidelity of $0.5$ when the channels are maximally noisy, i.e., with $p=0.5$. Note that this fidelity level is smaller than the input fidelity $F=0.6$. Interestingly, the performance of the LOCCNet architecture in Fig. \ref{fig: Ent_dist_LOCCNet} does not improve noticeably when optimized via the channel-aware criterion (\ref{eq: optimization expression}), as opposed to the noise-agnostic fidelity criterion considered in \cite{LOCCNet_Nature_2021}. In contrast, the proposed NA-LOCCNet with PQC architecture in Fig. \ref{fig: Ent_dist_proposed NA-LOCCNet} exhibits a significantly milder decrease in fidelity as $p$ grows, yielding the average output fidelity level of $F=0.8$ for $p=0.5$.

The advantages of NA-LOCCNet are further validated by Fig. \ref{fig: Ent_dist_Figure_output_fidelity_vs_input_fidelity.pdf}, which shows gains at all values of the input fidelity $F$. In particular, unlike the other schemes, NA-LOCCNet never yields an output fidelity lower than the input fidelity $F$.


\begin{figure}[htbp]
    \centering
    \includegraphics[height=2.7in]{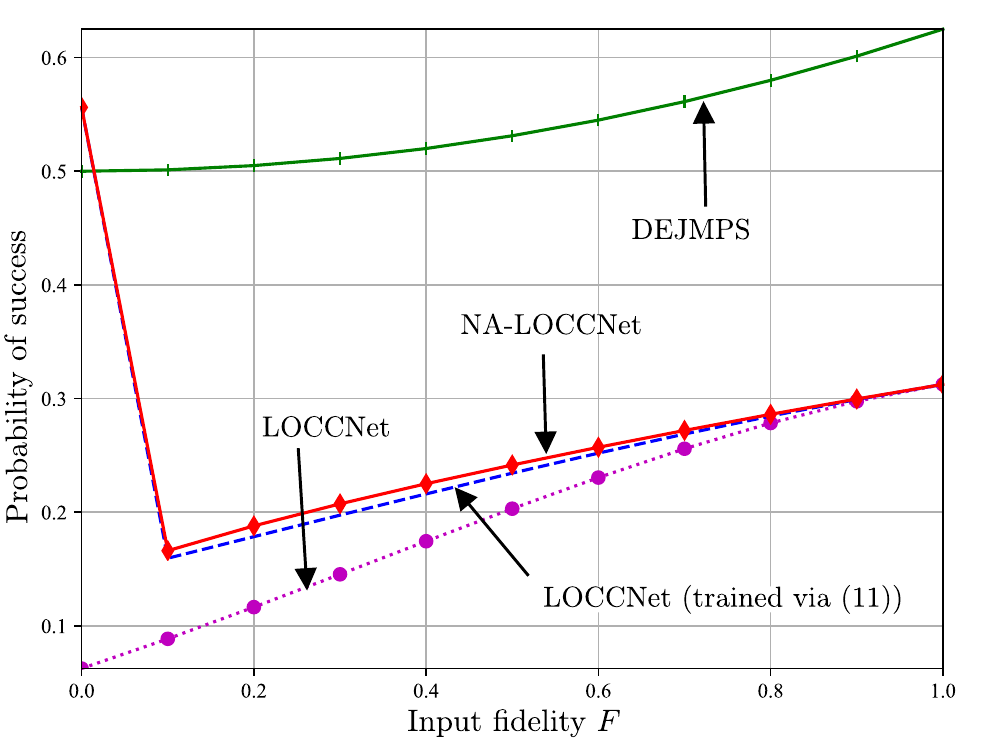}
    \caption{Probability of success as a function of the input fidelity $F$ in (\ref{eq: S-state}) for bit flip probability $p=0.25$ on the noisy classical channels from Alice and Bob to Charlie.}
    \label{fig: Ent_dist_Figure_prob_of_success_vs_input_fidelity.pdf}
\end{figure}


It is finally noted that the proposed approach, as well as LOCCNet \cite{LOCCNet_Nature_2021}, target the fidelity performance and not the probability of success. This point is illustrated in Fig. \ref{fig: Ent_dist_Figure_prob_of_success_vs_input_fidelity.pdf}, which shows the probability of success -- given by (\ref{eq: average success probability for 00 case}) for LOCCNet and NA-LOCCNet and by the sum of the probabilities for receiving the messages (0,0) and (1,1) at Charlie for DEJMPS -- as a function of the input fidelity $F$ for $p=0.25$. Overall, NA-LOCCNet is seen to offer a comparable probability of success as compared to LOCCNet, while improving the average fidelity.




\section{Learning Quantum State Discrimination with Noisy Classical Communication}
\label{section Learning Quantum State Discrimination}

In this section we first formulate the distributed quantum state discrimination problem and review the relevant state of the art protocols. We then propose NA-LOCCNet for distributed quantum state discrimination and give experimental results.

\subsection{Setting and performance metrics}
\label{subsection setting and performance metrics in QSD}

As in \cite{LOCCNet_Nature_2021}, we study the distributed quantum state discrimination problem illustrated in Fig. \ref{fig: QSD problem formulation}. In it, two agents, Alice and Bob, observe pairs of entangled qubits, and are tasked with detecting the joint quantum state of the qubit pairs. To this end, Alice and Bob can carry out local operations (LOs), as well as classical communication (CC) from Alice to Bob, i.e., they can implement an LOCC protocol. Unlike \cite{LOCCNet_Nature_2021}, we assume that the CC link between Alice and Bob is noisy. Applications of this setting include quantum sensor networks, as well as diagnostic functionalities for entanglement testing in the quantum internet \cite{QSD_applications,distributed_quantum_sensing,Book_Quantum_internet_second_quantum_revolution_2021}.

\subsubsection{Setting}
\label{subsubsection setting}

Alice and Bob share $S$ qubit pairs $(A_s,B_s)$ with $s\in [S-1]$, where each qubit $A_s$ is at Alice and each qubit $B_s$ is at Bob. Each qubit pair $(A_s,B_s)$ is entangled in one of two possible ways: The joint state of each pair $(A_s,B_s)$ is either given by the density matrix $\rho_0 = | \Phi^+ \rangle \langle \Phi^+ |$, with maximally entangled Bell state $| \Phi^+ \rangle = (|00 \rangle + |11 \rangle)/\sqrt{2}$; or it is in state $\rho_1 = \mathcal{N}(| \Phi^- \rangle \langle \Phi^- |)$, where $\mathcal{N}(\cdot)$ is an \textit{amplitude damping (AD)} channel and $| \Phi^- \rangle = (|00 \rangle - |11 \rangle)/\sqrt{2}$ is a maximally entangled Bell state orthogonal to $|\Phi^+\rangle$. The AD channel applies separately to the two qubits, and is expressed as 
\begin{equation}
    \mathcal{N}(\rho) = \sum_{i=0}^{1} \sum_{j=0}^{1} E_{ij} \rho E_{ij}^{\dagger},    
\end{equation}
where $E_{ij} = E_i\otimes E_j$ with Kraus matrices
\begin{equation}
    E_0 = |0 \rangle \langle 0 | + \sqrt{1-\gamma} |1 \rangle \langle 1 |
\end{equation}
and
\begin{equation}
    E_1 = \sqrt{\gamma} |0 \rangle \langle 1 |,
\end{equation}
where $0\leq \gamma \leq 1$ represents the noise parameter of the AD channel. For $\gamma=0$ the AD channel does not alter the input Bell state $| \Phi^- \rangle$, whereas for $\gamma=1$ the AD channel breaks the entanglement of the Bell state $| \Phi^- \rangle$, converting it to the product state $| 00 \rangle$. From \cite{LOCCNet_Nature_2021}, it is enough to consider the AD channel on a maximally entangled state, i.e., $|\Phi^- \rangle$, to make the two states, $\rho_0$ and $\rho_1$, non-orthogonal.
We note that results in this paper apply at a qualitative level to any other entanglement-breaking channel \cite{Book_Quantum_Information_Theory_wilde_2013}.

As seen in Fig. \ref{fig: QSD problem formulation}, Alice applies a parameterized quantum circuit (PQC) to the $S$ qubits $A_0,A_1,\cdots, A_{S-1}$ in her possession; then, it measures the $S$ qubits, and sends the $S$ classical bits obtained from the measurements to Bob. The PQC applied by Alice implements a $2^S \times 2^S$ unitary matrix $U^{A}(\theta^A)$ that is parameterized by vector $\theta^A$. Given that the input state for each qubit pair is $\rho_i$, with $i\in \{0,1\}$, the corresponding output state for the $2S$ qubits $A_0,A_1,\cdots, A_{S-1}$ and $B_0,B_1,\cdots, B_{S-1}$ can be written as 
\begin{equation}
    \label{eq rho^AB_i}
    \rho_i^{AB} = (U^A(\theta^A) \otimes I^B) \rho_{i}^{\otimes S} (U^A(\theta^A) \otimes I^B)^{\dagger},
\end{equation}
where $I^B$ is a $2^{S} \times 2^{S}$ identity matrix. The notation $\rho_{i}^{\otimes S}$ represents the state of the $S$ qubit pairs, with qubits ordered so that Alice qubits $A_0,A_1,\cdots,A_{S-1}$ are listed prior to Bob's qubits $B_0,B_1,\cdots,B_{S-1}$. 

Furthermore, Alice measures her qubits $A_0,A_1,\cdots,A_{S-1}$ using the $2^S$ projection matrices $\Pi_a^A = |a\rangle \langle a| \otimes I$ with $a\in \{0,1\}^S$, where $|a\rangle$ is the computational basis vector corresponding to bit string $a$. The measurement returns the output $a\in \{0,1\}^S$ with probability given by the Born rule, i.e., 

\begin{equation}
\label{eq prob of post measurement state at alice}
    P^{A}_{a|i} = \mathrm{tr}(\Pi_a^A \rho_{i}^{AB}).
\end{equation}

\noindent Note that the probability (\ref{eq prob of post measurement state at alice}) is conditioned on the true initial state $\rho_i$ of the qubit pairs. Alice communicates the $S$ classical bits $a \in \{0,1\}^S$ obtained from the measurement to Bob through a memoryless binary symmetric channel with bit-flip probability $p$.

As a result, Bob receives a message $\hat{a}\in \{0,1\}^S$ with probability $P_{\hat{a}|a}^{A\rightarrow B} = p^{d_{a,\hat{a}}} (1-p)^{S-d_{a,\hat{a}}}$, where $d_{a,\hat{a}}$ is the Hamming distance between the bit strings $a$ and $\hat{a}$. We note that this model can also account for measurement noise at Alice \cite{Measurement_noise_1_huggins2021efficient,Measurement_noise_2_Sharma_2020}. Therefore, the probability of receiving message $\hat{a}$ at Bob, when the qubit pairs initial state is $\rho_i$, is given by 
\begin{equation}
    P^{B}_{\hat{a}|i} =
    \sum\limits_{a\in \{0,1\}^S} P_{\hat{a}|a}^{A\rightarrow B}  P^{A}_{a|i},
\end{equation}
and the corresponding $2^S \times 2^S$ post-measurement density state of the $S$ qubits at Bob is
\begin{equation}
    \rho^B_{\hat{a}|i} = \sum\limits_{a\in \{0,1\}^S} P_{\hat{a}|a}^{A\rightarrow B} \frac{(\langle a| \otimes I) \rho_i^{AB} (|a \rangle \otimes I)}{P_{a|i}^A}.
\end{equation}
Depending on the message $\hat{a} \in \{0,1\}^S$ received at Bob, Bob performs a local operation given by the unitary $U^B(\theta^B_{\hat{a}})$, leaving the $S$ qubits in his possession in the density state

\begin{equation}
    \rho_i^B = \sum_{\hat{a} \in \{0,1\}^S} P^{B}_{\hat{a}|i}  U^B(\theta^B_{\hat{a}}) \rho^B_{\hat{a}|i} U^B(\theta^B_{\hat{a}})^{\dagger}.
\end{equation}

Finally, Bob applies a parity projective measurement on the $S$ qubits, using the projection matrices $\Pi_0^B = \sum_{\text{even }b}|b\rangle \langle b|$ and $\Pi_1^B = \sum_{\text{odd }b}|b\rangle \langle b|$, where ``even'' and ``odd" refer to the number of $1$'s in the bit string $b$, with $b\in \{0,1\}^S$. This produces the output $\hat{i}\in \{0,1\}$ with probability 
\begin{equation}
    P^B_{\hat{i}|i} = \mathrm{tr}(\Pi_{\hat{i}}^B \rho^B_{i}).
\end{equation}

One of the goals of this work is to design the PQC parameters $\theta^A$ and $\theta^B= \{\theta^B_{\hat{a}}\}_{\hat{a} \in \{0,1\}^S }$ at Alice and Bob such that the estimated state index $\hat{i} \in \{0,1\}$ at Bob equals the true state index $i$ with high probability. We specifically focus on protocols with single qubit pair, i.e., $S=1$ as studied in \cite{LOCCNet_Nature_2021}, in Section \ref{subsection LOCCNet in QSD}, and with two qubit pairs, i.e., $S=2$, in Section \ref{subsection Noise Aware-LOCCNet in QSD}.


\subsubsection{Performance metrics}
\label{subsubsection performance metrics in QSD}

Assuming that the two states $\rho_0$ and $\rho_1$ are selected a priori with equal probability, the average success probability is computed as
\begin{equation}
\label{eq avg succ prob}
    \mathrm{P}_{\hspace{-0.05cm} succ}(\theta^A, \theta^B) = \frac{1}{2} \sum_{i=0}^{1} P^B_{\hat{i}=i|i}.
\end{equation}
This probability is a function of the PQC parameters $\theta^A$ and $\theta^B$ at Alice and Bob respectively. We are interested in the problem of maximizing the average success probability

\begin{equation}
\label{eq optimization of prob of succ}
    \underset{\theta^A, \theta^B}{\max} ~ \mathrm{P}_{\hspace{-0.05cm} succ}(\theta^A, \theta^B).    
\end{equation}
Problem (\ref{eq optimization of prob of succ}) requires a search over the space of $|\theta^A| + \sum_{\hat{a} \in \{0,1\}^S}  |\theta^B_{\hat{a}}| $ PQC parameters, where $|\theta|$ represents the size of the vector $\theta$. This search can be carried out using standard optimization techniques, such as gradient descent.



We now discuss two upper bounds on the average success probability (\ref{eq avg succ prob}), namely the \textit{Helstrom bound} and the \textit{PPT bound}.

\subsubsection*{Helstrom bound}



Assume that all $S$ qubit pairs were available at a central node that could perform global measurements on all qubits. The maximum probability of successful detection in this system provides an upper bound on the probability of success for the distributed system under study. Allowing for a general \textit{positive operator valued measure (POVM)}, this approach yields the Helstrom bound \cite{Helstrom_1969,Helstrom_HOLEVO_1973}

\begin{equation}
\label{eq Helstrom bound}
    \mathrm{P}_{\hspace{-0.05cm} succ} \leq \frac{1}{2}+\frac{1}{4} \lVert \rho_0^{\otimes S} -\rho_1^{\otimes S} \rVert_{1},
\end{equation}
where $\lVert H \rVert_{1}$ represents the $l_1$-norm of the Hermitian matrix $H$, which is defined as the sum of the absolute values of the eigenvalues of matrix $H$.

\subsubsection*{Positive Partial Transpose (PPT) bound}

A tighter bound is obtained by restricting the type of measurements that are allowed at the central node having access to all $S$ qubit pairs. In particular, such restriction can be defined so as to include as a special case LOCC operations \cite{PPTPOVM_QSD_TIT2014}. The resulting PPT bound is obtained as the maximum value of the objective function of the semidefinite program (SDP) \cite{PPT_POVM_via_SDP_2013}
\begin{equation}
\label{eq PPT bound using PPT POVM SDP formulation}
\begin{aligned}
\max_{M_0,M_1} \quad & \frac{1}{2} \mathrm{tr}(M_0 \rho_0^{\otimes S} + M_1 \rho_1^{\otimes S})\\
\textrm{s.t.} \quad & \{M_i \succeq 0\}_{i=0}^{1}\\
  & \{M_{i}^{T_B} \succeq 0\}_{i=0}^{1}    \\
  & M_0 + M_1 =I ,\\
\end{aligned}
\end{equation}
where $M_{i}^{T_B}$ represents the partial transpose of the operator $M_{i}$ \cite{LOCCNet_Nature_2021,PPT_PAPER_W_Matthews_2008} with respect to the Hilbert space of Bob's qubits. We emphasize that Helstrom and PPT bounds does not depend on the communication between Alice and Bob as they assume centralized implementation.

\subsection{LOCCNet}
\label{subsection LOCCNet in QSD}


\begin{figure}[htbp]
    \centering
    \includegraphics[height=1.2in]{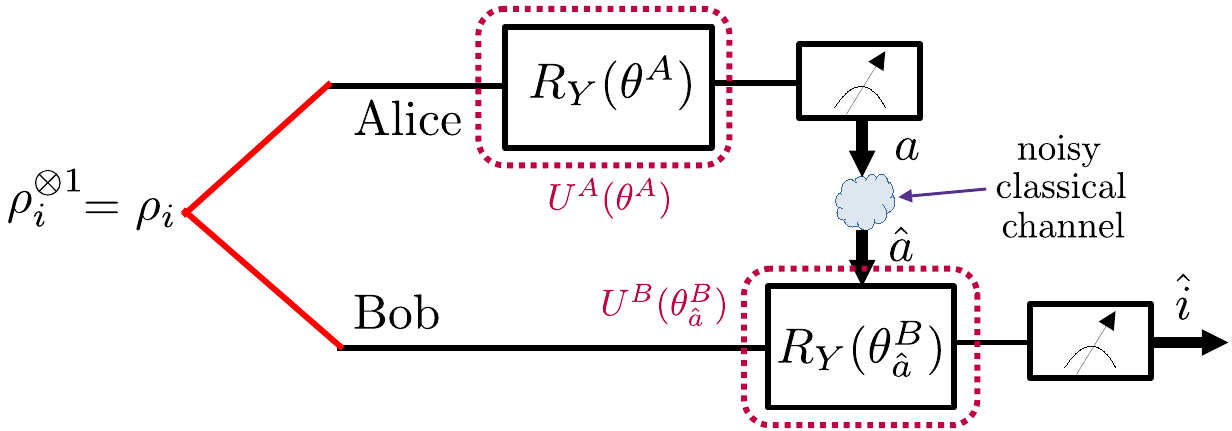}
    \caption{Illustration of the LOCCNet protocol \cite{LOCCNet_Nature_2021} for distributed quantum state discrimination, which operates on a single pair of qubits $(S=1)$.}
    \label{fig: QSD single pair}
\end{figure}


In this section, we review the LOCCNet protocol introduced in \cite{LOCCNet_Nature_2021}, which applies separately to each pair of qubits, i.e., $S=1$. As illustrated in Fig. \ref{fig: QSD single pair}, in LOCCNet the PQCs at Alice and Bob consists of Pauli $Y$-rotation gates, where the one-qubit Pauli $Y$-rotation gate is defined as \cite{Book_quantum_computing_quantum_information_Nielsen_chuang_2010}
\begin{equation}
    R_Y(\theta) = 
    \begin{bmatrix}
        \mathrm{cos~}{(\theta/2)} & -\mathrm{sin~}{(\theta/2)} \\
        \mathrm{sin~}{(\theta/2)} & \mathrm{cos~}{(\theta/2)}
    \end{bmatrix}.
\end{equation}
LOCCNet assumes a noiseless CC from Alice to Bob, and hence it addressed the special case of the optimization problem in (\ref{eq optimization of prob of succ}) with $p=0$. The optimized rotation angles are given as $\theta^A = \pi/2$ and $\theta^B_{\hat{a}} = (-1)^{\hat{a}} (\pi - \arctan{(\alpha)})$, with $\hat{a} \in \{0,1\}$ and $\alpha = (2-\gamma)/2$, where $\gamma \in [0,1]$ is the noise parameter of the AD channel.

In Section \ref{subsection experiments in QSD}, we will also evaluate the performance of the system illustrated in Fig. \ref{fig: QSD single pair} when the PQC parameters $\theta^A$ and $\theta^B$ are optimized by addressing problem (\ref{eq optimization of prob of succ}) with the correct value of the channel bit flip probability $p$.

\subsection{Noise Aware-LOCCNet}
\label{subsection Noise Aware-LOCCNet in QSD}


\begin{figure}[htbp]
    \centering
    \includegraphics[height=1.7in]{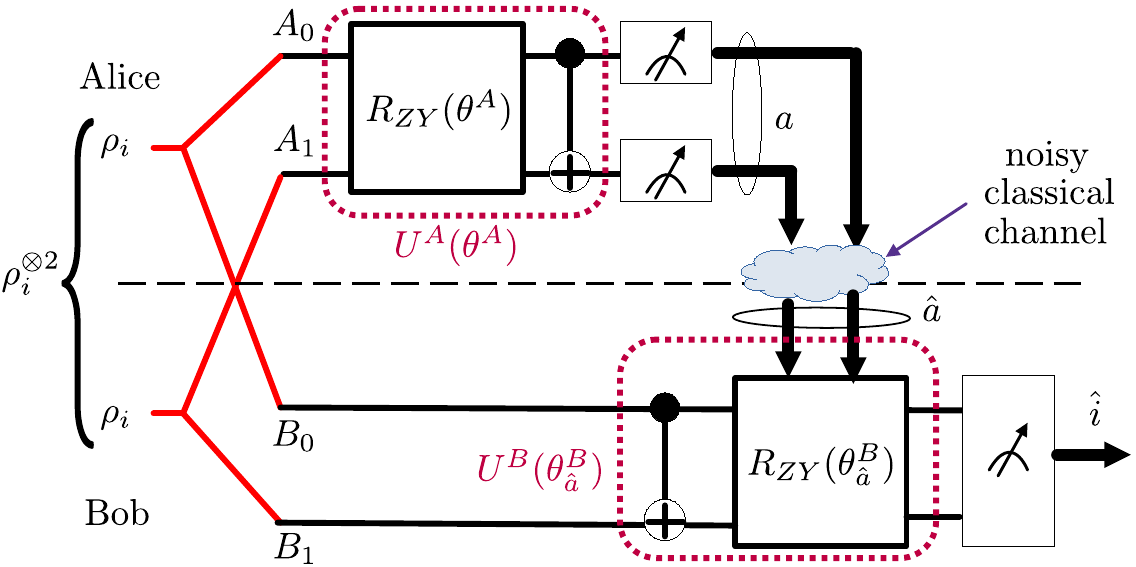}
    \caption{The proposed NA-LOCCNet protocol for distributed quantum state discrimination that operates over $S=2$ qubit pairs and adapts to the classical and quantum noise levels $p$ and $\gamma$.}
    \label{fig: QSD two pair}
\end{figure}


In this section, we introduce the NA-LOCCNet protocol, which operates on $S=2$ qubit pairs. There are two main innovations as compared to the LOCCNet protocol: $(i)$ We introduce an ansatz for the PQCs at Alice and Bob based on two-qubit rotation gates that can outperform the separate application of the LOCCNet protocol in Fig. \ref{fig: QSD single pair} to the two qubit pairs; $(ii)$ We propose the direct optimization of the noise-aware performance objective (\ref{eq optimization of prob of succ}), which is capable of adapting to the current classical noise level $p$, as well as to the quantum noise level $\gamma$.

For the PQCs, we adopt the architecture shown in Fig. \ref{fig: QSD two pair}, where the two qubit Pauli $ZY$-rotation gate is defined in (\ref{eq: two qubit ZY rotation gate}). Note that the Pauli $ZY$-rotation gates are followed at Alice, and preceded at Bob, by a controlled NOT (CNOT) gate. This ansatz has been selected through a partial numerical search. We  specifically explored other ansatzes with different two qubit and single qubit rotation gates, changing the position of CNOT gate before and after the rotation gates, and changing the control and target qubits of CNOT gates. The proposed ansatz in Fig. \ref{fig: QSD two pair} returned the best performance among the ansatzes that we considered.

For every value of the noise level $\gamma$ and bit flip probability $p$ we propose to optimize the average success probability in (\ref{eq avg succ prob}) over the rotation angles $\theta^A$ and $\theta^B_{\hat{a}}$ where $\hat{a} \in \{0,1\}^2$.

\subsection{Experiments}
\label{subsection experiments in QSD}

In this section, we evaluate the performance of the proposed NA-LOCCNet  protocols in the presence of a noisy CC link from Alice to Bob. We assume the availability of $S$ qubit pairs, and we consider LOCCNet, reviewed in Section \ref{subsection LOCCNet in QSD}, as the benchmark protocol.
As discussed in Section \ref{subsection Noise Aware-LOCCNet in QSD}, LOCCNet applies separately to the two qubit pairs, while the proposed NA-LOCCNet operates jointly on the two qubits pairs. LOCCNet is designed, for $S=1$, as in \cite{LOCCNet_Nature_2021} by setting $p=0$ in the optimization problem (\ref{eq optimization of prob of succ}), and we also evaluate the performance of the LOCCNet architecture in Fig. \ref{fig: QSD single pair} when the optimization is done by accounting for the actual value of $p$. We label this scheme as NA-LOCCNet $(S=1)$, since the design is noise aware. Optimization is done using Adam gradient descent optimizer \cite{Adam_optimzer_fundamental_paper}, with $0.01$ learning rate and $1000$ iterations. As performance bounds, we show the PPT bounds described in Section \ref{subsubsection performance metrics in QSD}, which are tighter than Helstrom bounds, for both the cases $S=1$ and $S=2$.


\begin{figure}[htbp]
    \centering
    \includegraphics[height=2.7in]{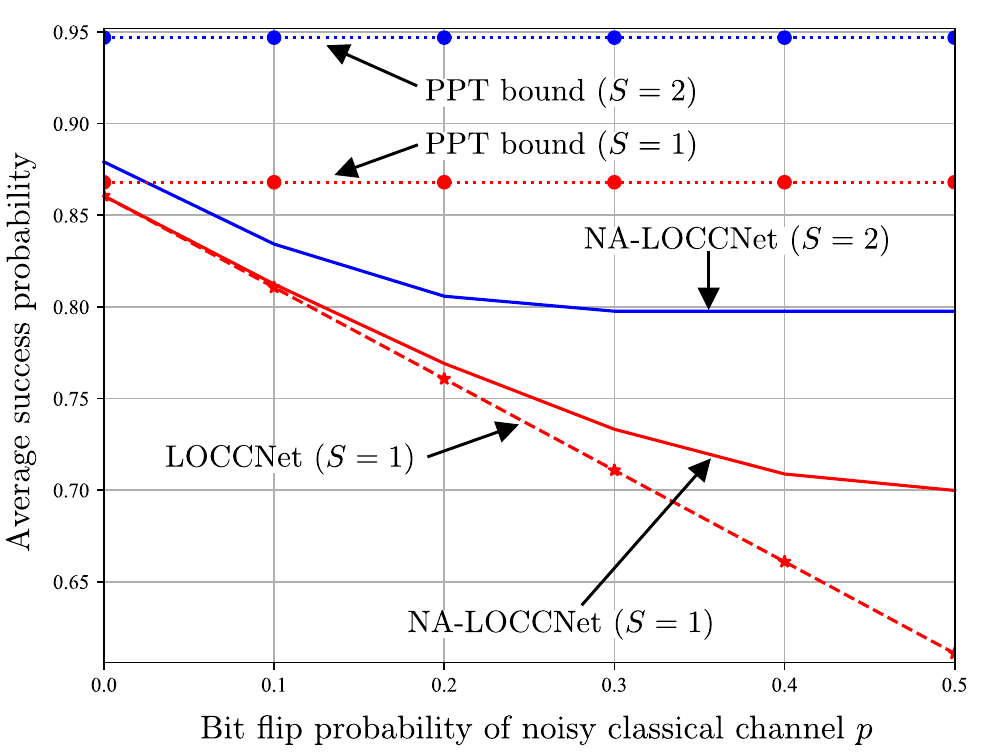}
    \caption{Average success probability as a function of the bit flip probability $p$ of the noisy classical channel from Alice to Bob for the AD channel noise parameter $\gamma = 0.8$.}
    \label{fig: avg succ prob vs prob of bit flip with gamma_0_point8}
\end{figure}



\begin{figure}[htbp]
    \centering
    \includegraphics[height=2.7in]{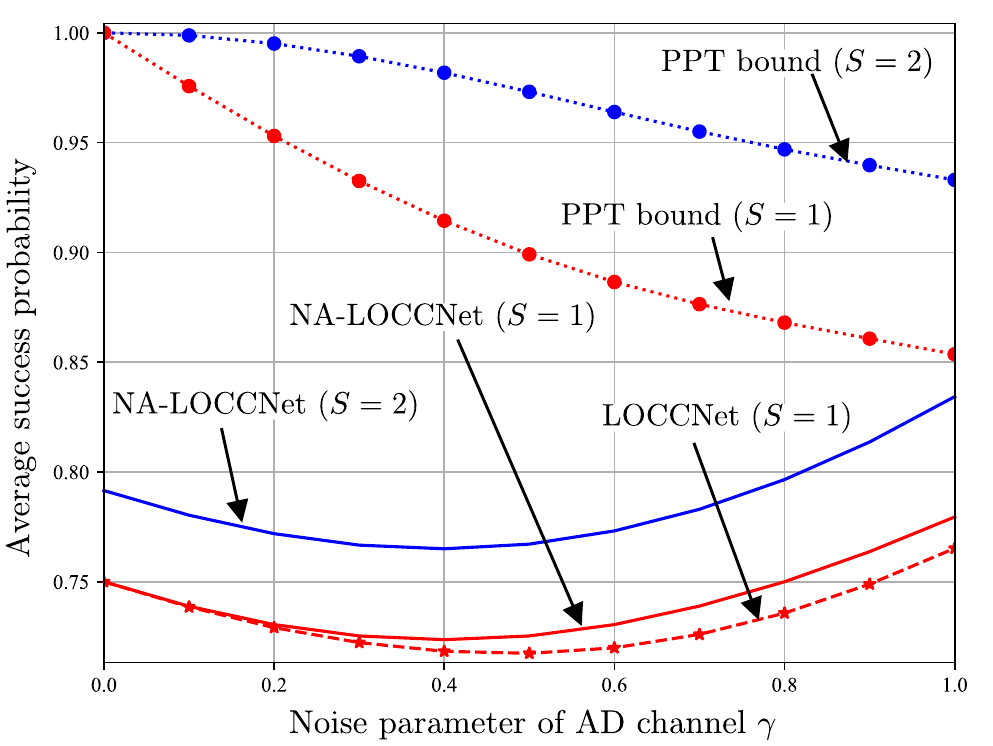}
    \caption{Average success probability as a function of the AD channel noise parameter $\gamma$ for the bit flip probability $p=0.25$ of the noisy classical channel from Alice to Bob.}
    \label{fig: avg succ prob vs gamma with p_0_point25}
\end{figure}

Fig. \ref{fig: avg succ prob vs prob of bit flip with gamma_0_point8} plots the average success probability (\ref{eq avg succ prob}) as a function of the bit flip probability $p$ of the noisy CC link by fixing the noise parameter of the AD channel to $\gamma=0.8$; while Fig. \ref{fig: avg succ prob vs gamma with p_0_point25} plots the same quantity as a function of the noise parameter of the AD channel $\gamma$ by fixing the bit flip probability of noisy CC to $p=0.25$. In both figures we use red lines for single-pair protocols, i.e., $S=1$, and blue lines for two-pair protocols, i.e., $S=2$. 


Fig. \ref{fig: avg succ prob vs prob of bit flip with gamma_0_point8} shows that, as the bit flip probability $p$ of noisy CC increases, the proposed NA-LOCCNet protocol vastly outperforms LOCCNet and NA-LOCCNet $(S=1)$. Specifically, the performance of LOCCNet reduces linearly as $p$ increases, whereas the proposed NA-LOCCNet is significantly more robust to communication noise. Note that, as suggested by comparing the PPT bounds with $S=1$ and $S=2$, the performance gain for $p=0.5$, i.e., for a completely noisy CC link, stems from the joint processing of two qubit pairs.


The advantages of NA-LOCCNet are further validated by Fig. \ref{fig: avg succ prob vs gamma with p_0_point25}, which demonstrates the gains of NA-LOCCNet at all values of the noise parameter of the AD channel $\gamma$. Interestingly, the probability of success first decreases and then increases as a function of the noise strength $\gamma$. To explain this behavior, consider the case $p=0.5$ of a fully noisy CC link and assume that Alice does not perform any operation on her qubits. In this case, Bob needs to distinguish $\rho_0^{\otimes 2}$ and $\rho_1^{\otimes 2}$ based solely on the local states $\mathrm{tr}_{A}(\rho^{\otimes 2}_0)$ and $\mathrm{tr}_{A}(\rho^{\otimes 2}_1)$, where $\mathrm{tr}_{A}(\cdot)$ represents the partial trace operation with respect to the qubits at Alice. The maximal probability of success for detection at Bob is given by the Helstrom bound (\ref{eq Helstrom bound}) as
\begin{equation}
    \label{eq prob of succ last equation}
    \mathrm{P}_{\hspace{-0.05cm} succ} = \frac{1}{2}+\frac{1}{4} \lVert \mathrm{tr}_{A}(\rho^{\otimes 2}_0) -\mathrm{tr}_{A}(\rho^{\otimes 2}_1) \rVert_{1}.
\end{equation}
The probability of success (\ref{eq prob of succ last equation}) takes the minimal value $0.5$ when there is no AD quantum noise, i.e., when $\gamma = 0$, since in this case we have $\mathrm{tr}_{A}(\rho^{\otimes 2}_0) = \mathrm{tr}_{A}(\rho^{\otimes 2}_1) = 0.5 I_4$. In contrast, at the other extreme, when $\gamma =1$, we have $\mathrm{tr}_{A}(\rho^{\otimes 2}_0) = 0.5 I_4$ and $\mathrm{tr}_{A}(\rho^{\otimes 2}_1) = |0\rangle \langle 0|$, and hence the probability of success (\ref{eq prob of succ last equation}) is given by $\mathrm{P}_{\hspace{-0.05cm} succ} = 0.75 > 0.5$. This argument suggests that, when the CC noise level $p$ is sufficiently large, the presence of an entanglement-breaking channel can be instrumental in improving the detection performance achievable via LOCC.

\section{Conclusions}
\label{section conclusions}

In this paper, we have studied the problems of distributed entanglement distillation and distributed quantum state discrimination in the presence of noisy classical communications. Specifically, we have proposed to train PQCs at the two parties so as to maximize the average fidelity in entanglement distillation and average success probability in quantum state discrimination. Simulation results have confirmed the advantages of the proposed NA-LOCCNet over the existing protocols designed for noiseless classical communications. Future work in entanglement distillation may involve the integration of the proposed scheme into a network protocol for entanglement distillation \cite{quantum_internet_protocol_stacl_review_2022}. 
For quantum state discrimination, it was observed that quantum entanglement-breaking noise on the observed system can be advantageous to improve the detection capacity when classical communication is noisy. Further increasing the number of qubit pairs $(S>2)$ may result in better protocols, and is a direction for future research.





\bibliographystyle{IEEEtran}
\bibliography{IEEEabrv,cite_Arxiv_final_version.bib}

\end{document}